\documentclass[preprint,12pt]{elsarticle} 

\makeatletter
\def\ps@pprintTitle{%
 \let\@oddhead\@empty
 \let\@evenhead\@empty
 \let\@oddfoot\@empty
 \let\@evenfoot\@empty
}
\makeatother

\usepackage{amssymb}
\usepackage[utf8]{inputenc} 
\usepackage[T1]{fontenc}    
\usepackage{hyperref}       
\usepackage{url}            
\usepackage{booktabs}       
\usepackage{multirow}
\usepackage{tabularx} 
\usepackage{makecell} 
\usepackage{caption}
\usepackage{subcaption} 
\usepackage{enumitem}
\usepackage{amsmath}
\usepackage{amsfonts}       
\usepackage{nicefrac}       
\usepackage{microtype}      
\usepackage{array}
\usepackage{fancyhdr}       
\usepackage{graphicx} 
\usepackage{threeparttable}
\usepackage{tikz}
\usepackage{xcolor}
\usepackage{ulem} 
\usepackage{tabularx} 
\usepackage{geometry}       
\usepackage{xcolor}
\setcellgapes{5pt} 

\begin{document}

\begin{frontmatter}
\title{Automated design for physics-informed modeling with convolutional neural networks}

\author[HKUSTGZ]{Wanyun Zhou}
\author[HKUSTGZ]{Haoze Song}
\author[HKUSTGZ]{Xiaowen Chu\corref{cor1}}
\cortext[cor1]{Corresponding author. Email address: xwchu@hkust-gz.edu.cn}

\address[HKUSTGZ]{The Hong Kong University of Science and Technology (Guangzhou), Guangzhou, 511453, P.R.China}

\begin{abstract}

Physics-informed convolutional neural networks (PICNNs) have emerged as a powerful extension of physics-informed neural networks (PINNs), offering superior generalization and efficiency for solving systems of partial differential equations (PDEs) modeling complex physical systems. However, current PICNNs, which rely on manual designs of both network architectures and loss functions tailored to specific PDEs, may fail to solve other types of PDEs that require different architectural designs and loss function formulations. To address these limitations, we propose leveraging automated machine learning (AutoML) to efficiently search for optimal network architectures and loss functions tailored to specific physical problems. By designing specialized search spaces and proposing a two-stage search strategy, our automated method substantially outperforms manually designed state-of-the-art models. It achieves up to a 59.8-fold reduction in prediction error and an average 13.31-fold error reduction across six diverse PDE systems spanning heat transfer, free fluid flow, and porous media flow, significantly enhancing the modeling capabilities for physical systems governed by steady or unsteady PDEs. An advantage of our automated method is that it can help researchers, without expertise in designing neural networks, develop the best models for their scientific problems.
\end{abstract}

\end{frontmatter}

\section{Introduction}
Partial differential equations (PDEs) play a crucial role in modeling a wide range of physical phenomena, including fluid dynamics \cite{patera1984spectral,wesseling2009principles}, heat transfer \cite{hady2012radiation,sierociuk2013modelling}, quantum mechanics \cite{laskin2002fractional,glimm2012quantum}, and electromagnetic wave propagation \cite{tsang2004scattering,rabczuk2019nonlocal}. These equations form the mathematical foundation for describing complex physical systems and their evolution 
over time and space. Solving these PDEs accurately and efficiently is therefore fundamental to numerous scientific and engineering applications. The traditional numerical methods for solving PDEs mainly rely on finite difference methods
and finite element methods \cite{1}. Both methods solve equations by discretizing the domain. Although numerical methods for solving differential equations can yield high-precision solutions, they necessitate very fine discretization of the domain, significantly increasing computational complexity and resulting in slower solving speed \cite{blendingoperator}. While utilizing coarse
grids for solving is faster, it results in lower accuracy \cite{2}. Moreover, nearly all numerical methods
require solutions to be obtained on the predefined grids, making it impossible to freely obtain
continuous values between grids \cite{3}. 

To address these limitations, deep learning approaches have emerged as promising alternatives for solving PDEs 
\cite{4, deepsurrogates,ranade2021discretizationnet,gasick2023isogeometric}. These methods can be broadly categorized into two main types: (1) data-driven approaches that learn the solution mapping directly from data \cite{rudy2017data,lu2021learning,maslyaev2020data,DBLP:conf/iclr/LiKALBSA21}, and (2) physics-informed neural networks (PINNs) that incorporate physical constraints during training \cite{5,18,jagtap2020conservative,NCscarce,computationalstructural,transferoperator}. While data-driven approaches have shown promising results, they require large amounts of high-quality training data generated from traditional numerical solvers, which can be computationally expensive to obtain. 
Additionally, these purely data-driven models may not guarantee physical consistency in their 
predictions \cite{wang2022hybrid}. To overcome the limitations of pure data-driven approaches while leveraging the advantages of deep learning, 
PINNs have become widely adopted. The fundamental concept behind PINNs is to exploit the expressive 
capabilities of deep neural networks to approximate the solution of a PDE while incorporating the 
governing equations as constraints in the loss function. This approach requires minimal or no training 
data as the physical laws guide the learning process, naturally enforces physical consistency in the 
predictions, and provides continuous solutions that can be evaluated at any point in the domain 
\cite{10}. Moreover, PINNs are particularly effective at handling irregular domains and complex 
boundary conditions that may be challenging for traditional numerical methods \cite{2,9}. \textcolor{black}{Recent efforts have introduced improvements to vanilla PINNs, such as hybrid PINNs \cite{fang2021high} which employ local fitting methods for more accurate differentiation, and hp-VPINNs \cite{kharazmi2021hp} which utilize more stable variational (weak) forms of PDEs for loss functions.}

However, most existing PINN frameworks including these recent improvements are based on Multi-Layer Perceptrons (MLPs).
Within this design, even minor modifications to the equations, including changes to physical parameters, boundary conditions, initial conditions, or source fields, require retraining the model from scratch, which can be highly inefficient \cite{11,12,multiresolution}. 
Consequently, such frameworks face significant challenges in practical scenarios, where physical conditions often change dynamically, limiting their applicability in real-world problems \cite{13,mechhoud2014estimation}. These challenges are particularly prominent for parametric PDEs which incorporate varying parameters in their equations with parameterized spaces encompassing diverse initial and boundary conditions, geometries, and source fields. Unlike non-parametric PDEs which are defined solely by variables and their derivatives without external varying parameters, each type of parametric PDE has multiple solution instances corresponding to different parameter configurations. When solving the equation under various parameter configurations, PINNs require complete retraining for each new parameter, making them highly inefficient in the parametric setting.
\textcolor{black}{To specifically address this challenge, methods like Isogeometric Neural Networks (IGN) \cite{gasick2023isogeometric} and Neural-Integrated Meshfree (NIM) methods \cite{du2024neural} have been explicitly designed to handle parametric PDEs without retraining. Both methods involve training a neural network to map the PDE parameters directly to the coefficients for a set of basis functions, which are then linearly combined to construct the final PDE solution. This allows them to learn solutions for a range of variables, such as varying boundary conditions or shape-defining parameters. However, the neural network component in both architectures remains a standard MLP, which cannot readily handle the extensive set of point-wise values that define a field, limiting their ability to process complex spatially-varying inputs such as source fields that are prevalent in the modeling of diverse physical systems.}

\textcolor{black}{To better address these challenges, convolutional architectures have shown great promise}. Physics-informed convolutional neural networks (PICNNs) \cite{11,12, 13, 14, 15, 16, 17} tackle these limitations by treating all parametric inputs of a PDE, from domain geometry and boundary conditions controlled by simple scalar coefficients to complex and spatially-varying source fields, as direct model inputs. This allows a single trained network to directly infer solutions for new parameters without retraining, achieving generalization across a family of related PDEs instead of being limited to a single instance \cite{12,13}.
For instance, Zhang et al. developed a PICNN for simulating transient two-phase Darcy flows in heterogeneous reservoir models with source/sink terms \cite{zhang2023physics}. Zhao et al. used PICNN to predict temperature fields with different heat source layouts \cite{13}. Rao et al. proposed a recurrent convolutional neural network for learning reaction-diffusion systems \cite{rao2023encoding}, while Liu et al. embedded discretized PDEs through convolutional residual networks in a multi-resolution setting for time-dependent PDEs \cite{multiresolution}. These architectural advantages of CNNs enable PICNNs to quickly adapt to variations in boundary or source fields, eliminating the need for extensive retraining required by MLP-based PINNs. This adaptability is particularly valuable for real-time or iterative simulations, such as those used in engineering, where physical conditions continuously evolve. Additionally, due to parameter sharing, PICNNs can handle large-scale problems more effectively than MLPs \cite{12,fuhg2023deep}, leading to better computational efficiency and scalability for practical applications.

In the context of PICNNs, each physical system is associated with a specific dataset that contains instances of a parametric PDE under various physical parameters manifested as different source fields, boundary conditions, and geometries, together with their corresponding high-resolution numerical solutions as ground truth solutions. 
Although PICNNs have been extensively applied in materials science, fluid mechanics, thermodynamics, 
geology, and other fields \cite{12, 17, 19, zhang2020physics,wandel2022spline}, several challenges remain: (1) Existing PICNNs are
manually designed, and many PDEs may not be solved accurately due to inappropriate network architectures. 
Since different physical problems require solving various types of PDEs, each potentially favoring different
network structures, finding the optimal PICNN architecture for each PDE through numerous manual experiments 
is impractical. (2) For different PDEs, the computational factors of the loss functions vary across these 
datasets \cite{kharazmi2021hp,du2024neural}. Current PICNNs mainly use vanilla loss functions, often neglecting the weighting strategies for 
different residual points, a technique frequently used in PINNs \cite{1,30,39}. However, even in PINNs, weighting 
strategies vary widely across different PDE datasets, and those effective for one PDE dataset may significantly 
increase prediction error on others compared to the vanilla loss function. (3) Existing PICNNs are
limited to solving steady-state PDEs and cannot directly address more complex spatiotemporal PDEs, which require 
modeling both spatial and temporal dynamics. Regarding the first challenge, only a few studies have applied automatic machine learning (AutoML) to design PINN models \cite{23, 24, 25,40}. However, these studies focus exclusively on MLP-based PINNs and do not explore the use of AutoML with CNNs as the underlying framework. The reliance on MLPs makes these methods applicable only to certain PDEs and requires retraining when parameters in the equations change. Furthermore, in AutoML for such PDEs, the search objective is limited to selecting a PDE-driven loss, which is shown to be inappropriate in the following section. As for the second challenge, the automatic computation of appropriate loss functions in both PINNs and PICNNs remains unexplored. 

Given these limitations in existing approaches and to address the first and second challenges, we propose using AutoML to search for optimal loss functions and network architectures for PICNNs. While AutoML has shown promise in various domains \cite{31,elsken2019neural,hutter2019automated,he2018amc}, its application to physics-informed modeling is not straightforward. Unlike general-purpose AutoML frameworks, a physics-informed approach requires carefully designed search objectives that balance physical consistency with prediction accuracy, as well as tailored search spaces for loss functions and network architectures. These search spaces must incorporate physical constraints, innovative weighting strategies, and the ability to efficiently explore architectures that capture complex physical dynamics. Poor design choices in any of these aspects can lead to significant prediction errors. Given these technical complexities, the manual design of effective PICNNs remains a significant barrier for physicists, mathematicians, engineers, and other domain practitioners. Our automated framework alleviates these challenges by enabling researchers who lack
expertise in neural network design to discover optimal models and loss functions for their specific problems. To further address the third challenge of solving spatiotemporal PDEs with PICNNs, inspired by \cite{16}, we embed a ConvLSTM module \cite{shi2015convolutional} into our constructed 
search space to capture temporal dynamics while preserving PICNNs' spatial modeling capabilities, enabling 
the model to directly infer solutions for different initial conditions without requiring retraining for each change.

Building upon these advancements, our framework involves a systematic process: Firstly, we delve into the correlation between the PDE-driven loss and prediction errors within the PICNNs to determine the best search objective. Secondly, by considering factors that may influence the loss function, we construct an operator-infused loss function search space that encompasses these factors and includes several residual adjustment operations, and then utilize hyperparameter optimization to explore an appropriate loss function for each dataset. Concurrently, we define the fundamental architecture of PICNNs across different datasets and establish an appropriate search space. This involves choosing the entire-structured search space instead of the cell-based search space as the overall network structure of PICNNs. Finally, we employ physics-informed neural architecture search (NAS), which utilizes policy-based reinforcement learning specifically tailored for PICNNs to search for the best network architecture within the search space. 

To the best of our knowledge, we are the first to propose an automated framework that enables researchers to automatically discover optimal PICNN architectures and loss functions for various physical systems, effectively addressing manual design challenges while empowering physicists, mathematicians, engineers, and others lacking expertise in neural network design to develop the best models for their scientific problems. Our contributions encompass both methodological innovations and practical advances. We introduce an operator-infused loss function search space that incorporates physics-aware components and innovative weighting strategies, leading to more accurate predictions across different physical systems. Additionally, we design specialized search spaces for the network architectures of PICNNs, which enhances their ability to handle both spatial and temporal changes in physical systems, making the search process more comprehensive and effective. By leveraging this AutoML framework, our method significantly surpasses manually designed state-of-the-art models in solving PDEs, achieving up to a 59.8-fold reduction in prediction error and an average of 13.31-fold error reduction across six diverse PDE systems spanning heat transfer, free fluid flow, and porous media flow.

\section{Results}
\subsection{Overall framework of Auto-PICNN}
\begin{figure}[t]
    \centering
    \includegraphics[width=0.95\linewidth]{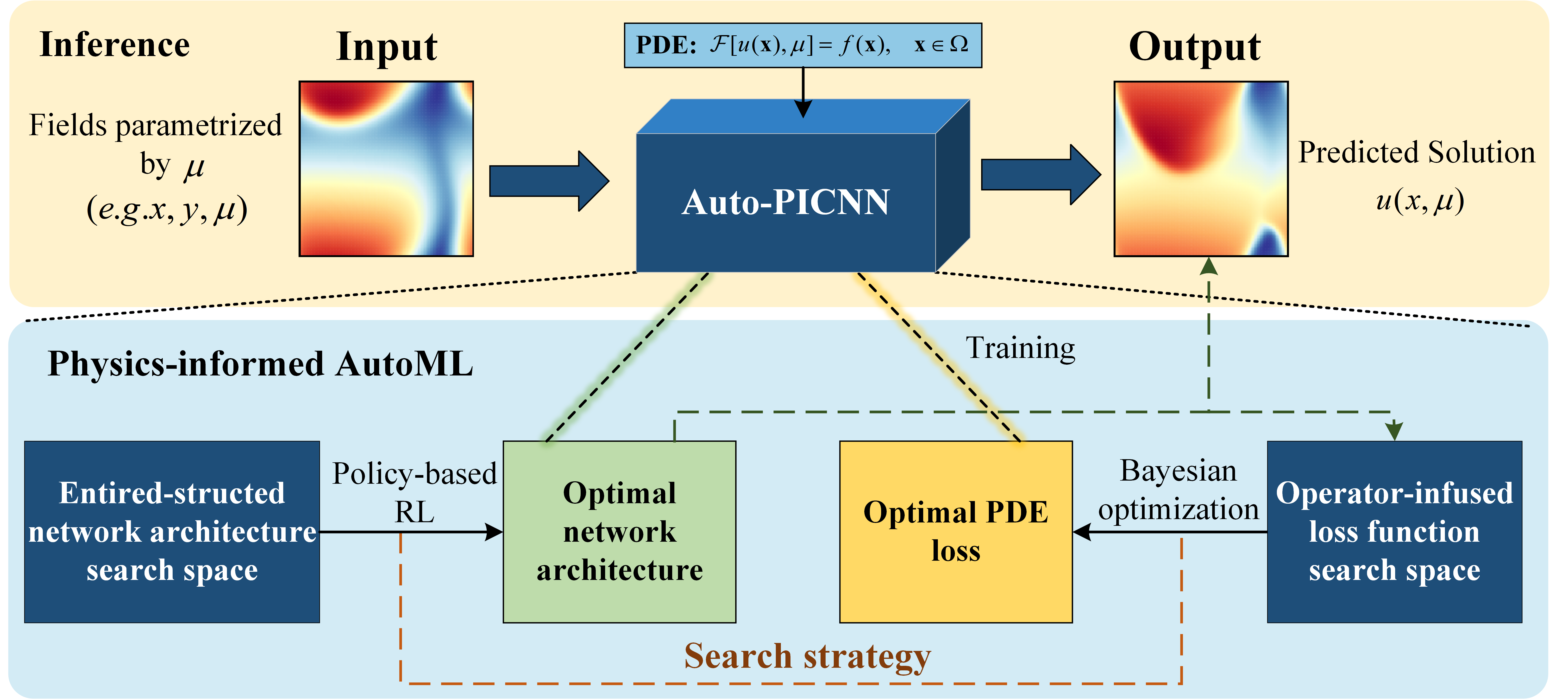}
    \caption{\textbf{Schematic illustration of the automated framework for physics-informed convolutional neural network design.} The framework leverages automated machine learning (AutoML) with a two-stage search strategy to identify optimal network architectures and loss functions for training. Once trained, the model can directly infer solutions for parametric PDEs with any unseen parameter configurations without requiring retraining.}
    \vspace{-0.2cm}
    \label{framework}
\end{figure}

Auto-PICNN is an advanced and automated framework that streamlines and enhances the development of physics-informed modeling for solving a wide array of parametric PDEs. The framework leverages physics-informed AutoML, in which the essential components (search objective, search space construction, and search strategy selection) are tailored with domain-specific considerations for PDE solving, enabling automatic design and optimization of both loss functions and network architectures in PICNNs. The overall structure and workflow of Auto-PICNN, as shown in Figure \ref{framework}, consists of the following steps:

1. Data preparation: For the training and evaluation of Auto-PICNN, we prepare a dataset for each parametric PDE consisting of multiple instances with varying parameter settings and their corresponding high-resolution numerical solutions. The model is trained using PDE-driven loss functions to ensure physical consistency, while a small subset of samples is selected from the training set as the validation set to evaluate the prediction error during the training phase. Once trained, as shown in the inference stage of Figure \ref{framework}, the Auto-PICNN model can efficiently handle new parameter configurations: for any unseen fields parameterized by $\mu$, it can instantly generate solutions through direct inference without requiring retraining.

2. Define search objectives: To guide the optimization process, we need to determine appropriate search objectives. Instead of relying on commonly used PDE-driven losses, our experimental results (see 2.2 Search objective) show that the relative $L^{2}$ error on the validation set serves as a more suitable search objective as it ensures that the prediction error across different instances of a PDE family remains relatively uniform.

3. Design the loss function search space: To construct effective PDE-driven loss functions for training PICNNs, Auto-PICNN introduces an innovative operator-infused loss function search space. This search space is composed of operators such as unary operators, gradient enhancement operators, and residual-based operators representing point-wise weighting strategies. These operators are tailored to better capture the unique characteristics of PDEs and the physical systems they represent. The search space also incorporates various physics-aware components that are critical for the effectiveness of the loss function, including constraints on boundary conditions and convolution kernels for calculating spatial derivatives. Together, these elements enable the search space to incorporate multiple new strategies for effective loss formulation, achieving better alignment with the physical properties of PDEs. 

4. Design the neural network architecture search space: Concurrently, Auto-PICNN designs a search space for exploring optimal neural network architectures. The framework adopts an entire-structured search space for PICNN architectures, replacing the cell-based search spaces commonly used in AutoML. When handling spatiotemporal PDEs, this search space seamlessly integrates ConvLSTM \cite{shi2015convolutional} modules, enabling PICNNs to efficiently capture both spatial and temporal dynamics.  

5. Employ a two-stage search strategy: With these two search spaces in place, Auto-PICNN employs a two-stage search strategy. In the first stage, Bayesian optimization is used to navigate the operator-infused loss function search space and identify the most effective loss function. In the second stage, a policy-based reinforcement learning approach is employed for neural architecture search (NAS). A Long Short-Term Memory (LSTM)-based controller generates candidate architectures, which are iteratively refined to discover the best-performing network architecture.  

For technical implementation of the Auto-PICNN framework, see the Methods section for details.

\begin{figure}[t]
    \centering
    \includegraphics[width=0.95\linewidth]{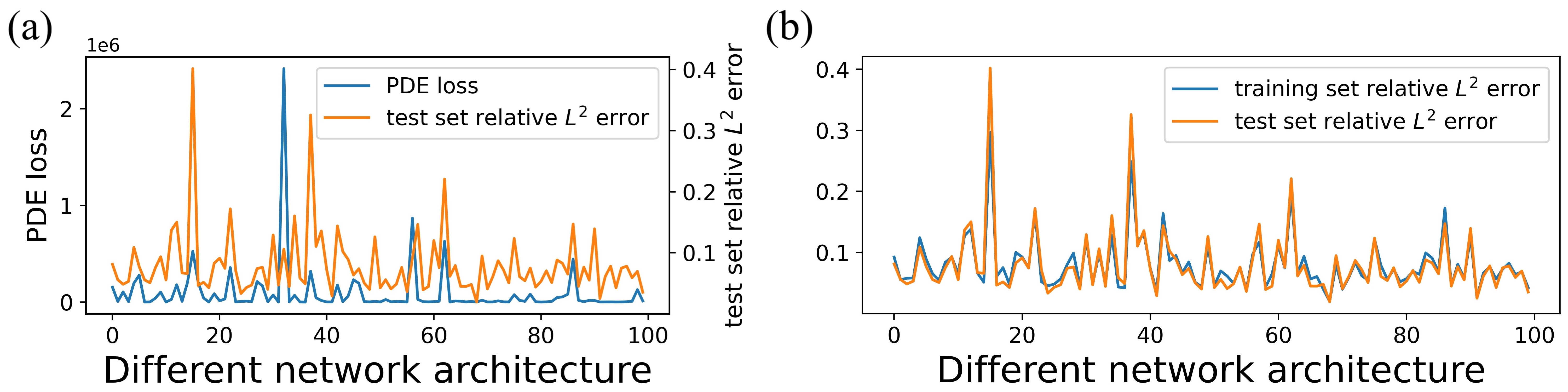} 
    \caption{\textbf{Comparison of physics-driven loss and prediction error across different network architectures.} Results are obtained from 100 randomly selected network architectures trained on the parametric heat equation. (a) presents the PDE-driven loss on the training set and the relative $L^{2}$ error on the test set, while (b) compares the relative $L^{2}$ errors on the training and test sets.}
    \label{fig:main_figure}
\end{figure}

\subsection{Search objective}
When dealing with non-parametric PDEs which are defined solely by the variables and their derivatives, without external parameters, we can only choose the PDE-driven loss $L(\theta)$ as the search objective to determine the loss function and architecture. However, for PICNNs, the PDE-driven loss and relative prediction error are not uniform across different architectures. 
\textcolor{black}{Taking the parametric heat equation with 7 samples representing different boundary temperatures as an example, we designed the experiment by randomly selecting 100 different network architectures and trained each architecture with the same training set. To ensure robustness across different data distributions, we systematically tested 6 different train-test splits ranging from 1 training sample to 6 training samples.
For each architecture, we recorded the PDE-driven loss on the training set and the relative $L^{2}$ error on both training and test sets. The relative $L^{2}$ error quantifies the mean squared difference between the network's predicted solutions and the ground truth, normalized by the $L^{2}$ norm of the ground truth, which ensures that the metric is scale-invariant.}
As shown in Figure \ref{fig:main_figure} (a), it can be observed that when the PDE-driven loss is high, the relative $L^{2}$ error can be low, and vice versa (\textcolor{black}{results from other data splits are provided in Supplementary Note 1}). The discrepancy observed may be attributed to the phenomenon that using some neural network architectures to solve PDEs may fall into the trivial solutions. In these cases, the training of PICNNs minimizes PDE-driven loss by adopting simplistic solutions that, while adhering to some basic PDE constraints, fail to capture the full complexity of the physical problem, leading to higher relative $L^{2}$ errors. Such outcomes highlight the fact that the PDE-driven loss alone may not be a reliable indication of the actual prediction performance, which suggests that it is difficult for non-parametric PDEs to find an appropriate search objective. In contrast, for parametric PDEs, we can split its dataset into the training and the test set. As shown in Figure \ref{fig:main_figure} (b), the prediction error of the training set and the test set is relatively uniform. In addition, selecting a small number of samples from the training set as the validation set with its prediction error being the search objective ensures the effectiveness of the search while maintaining the advantage of PICNNs in saving time and resources without generating labels on a large number of data points.

\subsection{Benchmarks and Baselines}
To ensure the robustness and wide applicability of the findings across different scenarios in the PICNNs, we select a diverse range of datasets in previous studies which includes heat equations with different boundary conditions, Poisson equations with varying source fields, Poisson equations with different heat source layouts, Darcy flow with different input property fields, Navier-Stokes equations with vary-ing geometries, and 2D Burgers' equations as benchmarks. The detailed introduction of the datasets can be seen in Supplementary Note 2.

To comprehensively assess the performance of Auto-PICNN, we employ three different baselines to evaluate its effectiveness:

\textbf{1. Manually designed baseline models}. The models of PICNNs (including PhyGeoNet \cite{12}, PI-UNet \cite{13}, and PDE-surrogate \cite{11}, PhyCRNet \cite{16}) corresponding to the six datasets mentioned above were selected as the manually designed baseline models. These baselines have demonstrated strong performance for their respective PDE datasets, with further details provided in Supplementary Note 4.

\textbf{2.Vanilla MSE loss baseline}. To demonstrate the effectiveness of using AutoML in loss function optimization, we construct a vanilla MSE baseline used in most PICNNs. This serves as the default loss function for training the PICNNs, allowing us to measure how much AutoML can improve the effectiveness of the loss function.

\textbf{3. Cell-based search space baseline}. To demonstrate the effectiveness of our entire-structured search space for network architecture, we also establish a cell-based search space baseline. This baseline allows for a comparison that highlights the benefits of our chosen entire-structured approach in optimizing network architectures.

\subsection{Comparison with the baseline models}
The manually designed baseline models corresponding to the six datasets were compared with the models obtained through our automated machine learning methods (including the search for loss function and network architecture). For each dataset, both the baseline model and the search-based model were trained and tested on the same set.

\begin{figure}[t]
    \centering
    \includegraphics[width=0.98\linewidth]{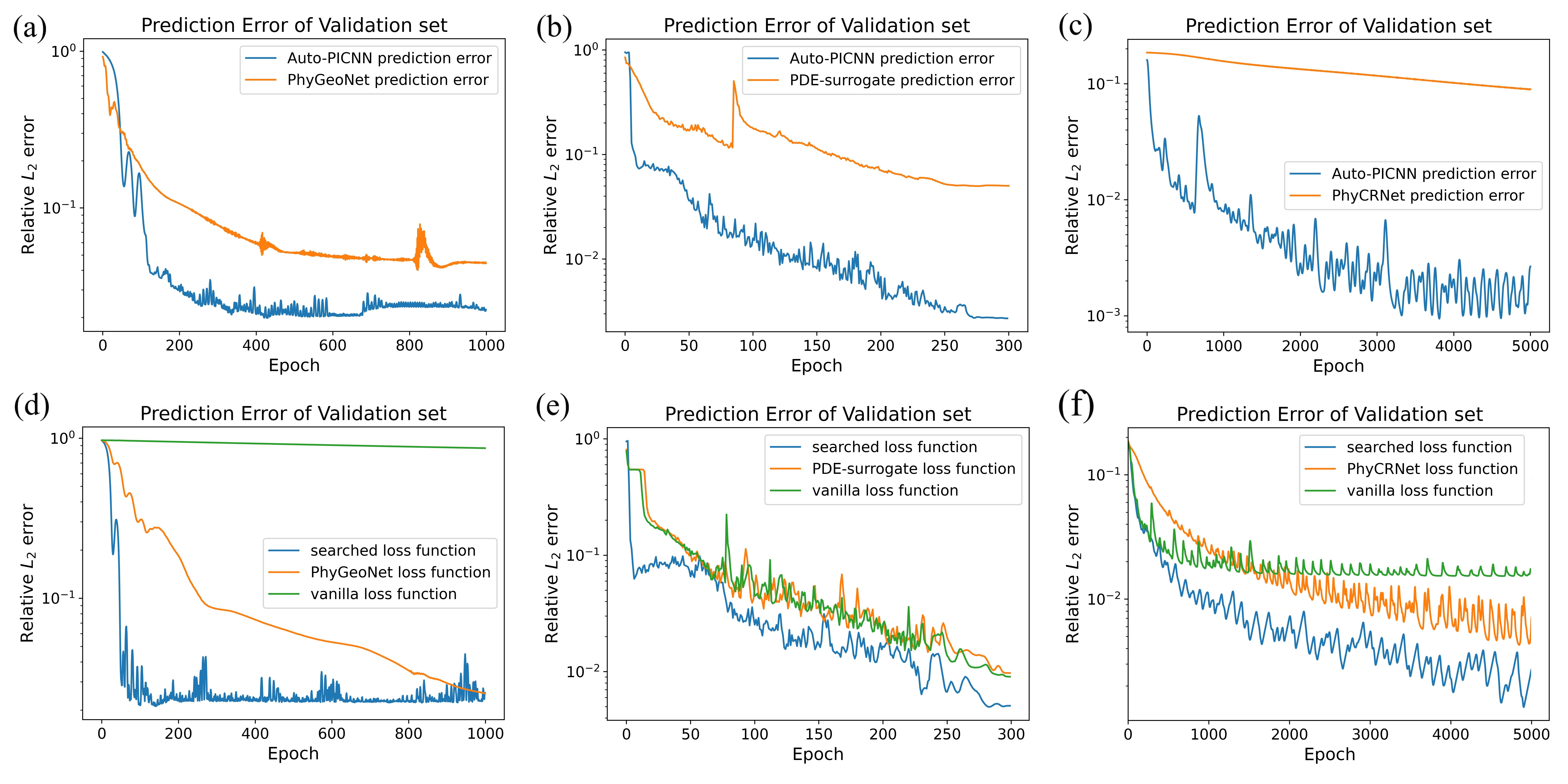}
    \caption{\textbf{Prediction error curves on the validation set during training for baseline models and our model on different partial differential equation datasets.} (a), (b), and (c) show the relative $L^{2}$ error of our model termed Auto-PICNN compared to the baseline models PhyGeoNet, PDE-surrogate, and PhyCRNet for the heat equations, Darcy flow, and 2D Burgers' equations, respectively. (d), (e), and (f) compare the relative $L^{2}$ error obtained using three different loss functions including our searched loss function, baseline models' loss function, and vanilla MSE loss for the heat equations, Darcy flow, and 2D Burgers’ equations, respectively.}
    \label{hblosscurve}
\end{figure}

For heat equations with different boundary conditions, Darcy flow with varying input property fields, and 2D Burgers’ equations, Figure \ref{hblosscurve} (a), (b), and (c) show the training curves of the relative prediction error for the baseline models and Auto-PICNN on the validation set. In all cases, Auto-PICNN demonstrates significantly faster convergence and lower final prediction errors compared to the baseline models. For instance, in Figure \ref{hblosscurve} (c), the baseline PhyCRNet struggles to reduce the relative $L^2$ error below $10^{-1}$, while Auto-PICNN achieves an error on the order of $10^{-3}$—nearly two orders of magnitude smaller. This highlights Auto-PICNN's superior accuracy and efficiency in physic-informed modeling. 
Figure \ref{combined_figure} (a), (b), and (c) also visualize the predicted solutions and point-wise prediction errors of Auto-PICNN compared to the baseline models for a randomly selected sample on the test set across different PDE datasets. For Darcy flow and 2D Burgers’ equations, Auto-PICNN effectively reduces significant errors in regions with comparatively sharper spatial transitions, a challenging characteristic associated with the stiffness of PDEs. For the heat equation on an annular domain, Auto-PICNN accurately captures heat transfer across the entire region, including the inner circular area, where baseline models struggle due to the difficulty of uniformly propagating the known boundary conditions on both sides of the annulus. These results demonstrate Auto-PICNN’s robustness in handling complex physical and geometric scenarios. The training curves and predicted solutions for other PDE datasets, as well as additional comparisons between Auto-PICNN and other baseline models, can be found in Supplementary Note 5.

\begin{table}[t]
\centering
\resizebox{\columnwidth}{!}{
\begin{tabular}{c|c|cc}
\hline
Dataset & Model & \multicolumn{2}{c}{Error} \\ \hline\hline
\multirow{2}{*}{\shortstack{Heat equations with different boundary conditions}} & PhyGeoNet & \multicolumn{2}{c}{0.0523} \\ \cline{2-4}  
 & \textbf{Auto-PICNN} & \multicolumn{2}{c}{\textbf{0.0177}} \\ \hline
\multirow{2}{*}{\shortstack{Poisson equations with varying source fields}} & PhyGeoNet & \multicolumn{2}{c}{0.0744} \\ \cline{2-4}  
 & \textbf{Auto-PICNN} & \multicolumn{2}{c}{\textbf{0.0131}} \\ \hline
\multirow{2}{*}{\shortstack{Poisson equations with different heat source layouts}} & PI-UNet & \multicolumn{2}{c}{0.0271} \\ \cline{2-4}  
 & \textbf{Auto-PICNN} & \multicolumn{2}{c}{\textbf{0.0162}} \\ \hline
\multirow{2}{*}{\shortstack{Darcy flow with different input property fields}}
 & PDE-surrogate  & \multicolumn{2}{c}{0.028} \\ \cline{2-4} 
 & \textbf{Auto-PICNN} & \multicolumn{2}{c}{\textbf{0.0032}} \\ \hline
\multirow{2}{*}{\shortstack{2D Burgers' equations with initial conditions}}
 & PhyCRNet  & \multicolumn{2}{c}{0.1847} \\ \cline{2-4} 
 & \textbf{Auto-PICNN} & \multicolumn{2}{c}{\textbf{0.0031}} \\ \hline
\multirow{2}{*}{\shortstack{Navier-Stokes equations with varying geometries}} & PhyGeoNets & \multicolumn{1}{c|}{0.0797} & 0.3543 \\ \cline{2-4} 
 & \textbf{Auto-PICNN} & \multicolumn{1}{c|}{\textbf{0.0564}} & \textbf{0.2966} \\ \hline
\end{tabular}
}
\caption{\textbf{Test set errors for the baseline models and our Auto-PICNN across six PDE datasets.} The reported error is MAE for the Poisson equations with different heat source layouts and relative $L^2$ error for the other datasets. For the Navier–Stokes equations with varying geometries, the first error column corresponds to velocity and the second to pressure. The best results are highlighted in bold.}

\label{table1}
\end{table}

For the test sets, the numerical results for all PDE datasets are shown in Table \ref{table1}. Experimental findings indicate that our searching method significantly surpasses the manually designed baseline models in solving parametric PDEs. The prediction error is reduced by factors of 2.72, 5.68, 1.67, 8.75, 1.30 and 59.78 across the six datasets, with an average reduction by a factor of 13.31. The superiority of Auto-PICNN on various datasets also ensures that it can be applied to solve a wide range of PDEs with different characteristics. 

\begin{figure}[!htb]
    \centering
    \includegraphics[width=0.95\linewidth]{Figure/combined_main_solution.jpg}
    \caption{\textbf{Visualizations of the predicted solutions and point-wise prediction errors of our model and baseline models on different partial differential equation datasets.} (a), (b), and (c) visualize the predicted solutions and their corresponding point-wise errors of our model termed Auto-PICNN and baseline models PhyGeoNet, PDE-surrogate, and PhyCRNet on the heat equations, Darcy flow, and 2D Burgers' equations, respectively. (d), (e), and (f) visualize the predicted solutions and point-wise errors obtained using different loss functions including our searched loss function, baseline models' loss function, and vanilla MSE loss for the heat equations, Darcy flow, and 2D Burgers' equations, respectively.
    For the ground truth and prediction visualizations, colorbars indicate the physical field's magnitude, representing temperature for heat equations, pressure for Darcy flow, and fluid velocity for Burgers' equations. In the error visualizations, colorbars represent the point-wise absolute error between predictions and ground truth.}
    \label{combined_figure}
\end{figure}

\subsection{Loss function comparison}
To assess the performance improvements achieved by our automatic loss function searching method, we compare the loss function after searching with the vanilla MSE loss baseline. The prediction error is reduced by factors of 30.66, 1.10, 6.30, 2.24, 1.25 and 12.53 across the six PDE datasets, with an average reduction by a factor of 9.01. These significant improvements demonstrate the effectiveness of the automatically searched loss functions in adapting to different PDE datasets.

We also compare it with the loss functions designed by the baseline models. Taking the heat equations with different boundary conditions, Darcy flow and 2D Burgers’ equations as examples, PhyGeoNet, PDE-surrogate and PhyCRNet are chosen as the manually designed baseline models respectively. Figure \ref{hblosscurve} (d),(e) and (f) show that the loss function after searching significantly reduces the relative $L^{2}$ error on the validation set compared to the loss function of the baseline model and the vanilla MSE loss. This can also be illustrated by visualising the predicted solutions and their point-wise errors against the ground truth (see Figure \ref{combined_figure} (d), (e), and (f)). Taking the heat equations with different boundary conditions as an example, Figure \ref{hblosscurve} (d) demonstrates that the vanilla MSE loss function fails to converge, resulting in persistent high errors and severe artifacts during training. Specifically, as shown in Figure \ref{combined_figure} (d), large portions of the domain exhibit nearly uniform predictions, highlighting the tendency to fall into a trivial solution due to the inappropriate choice of loss function. In contrast, the searched loss function successfully approximates the true solution, addressing these challenges effectively.

The specific values of the prediction errors in solving PDEs on all six datasets under different loss functions as well as the training curve and the predicted solution of other PDE datasets can be seen in Supplementary Note 6. For the Poisson equation with different heat source layouts, PI-UNet \cite{13} used MAE as the evaluation metric. To ensure a fair comparison, we use the same evaluation metrics for analysis and comparison. 

The search results for the components and operations of the loss function can be found in the Supplementary Note 7, which highlights that different PDE datasets favor different loss functions. For instance, datasets such as heat equations with different boundary conditions benefit from gradient-enhanced operators and the use of pointwise loss gradients to compute the residual matrix. In contrast, datasets like Poisson equations with varying source fields prefer simpler configurations without these enhancements.

\begin{table*}[htbp]
\centering
\resizebox{0.96\textwidth}{!}
{ 
\begin{tabular}{c|c|c}
\hline
\rule{0pt}{1.2em}Dataset & \rule{0pt}{1.2em}Search strategy & \rule{0pt}{1.2em}Relative ${L^{2}}$ error \\ \hline \hline
\multirow{3}{*}{Heat equations with different boundary conditions} & Reinforcement learning & \rule{0pt}{1.2em}\textbf{0.0163} \\ \cline{2-3} 
 & Random Search & \rule{0pt}{1.2em}0.0426 \\ \cline{2-3} 
 & DARTS & \rule{0pt}{1.2em}0.0740 \\ \cline{2-3} 
 & ENAS & \rule{0pt}{1.2em}0.0602 \\ \hline
 \multirow{3}{*}{Poisson equations with varying source fields} & Reinforcement learning & \rule{0pt}{1.2em}\textbf{0.0120} \\ \cline{2-3} 
  & Random Search & \rule{0pt}{1.2em}0.0123 \\ \cline{2-3} 
 & DARTS & \rule{0pt}{1.2em}0.0121 \\ \cline{2-3} 
 & ENAS & \rule{0pt}{1.2em}0.0121 \\ \hline
\multirow{3}{*}{Poisson equations with different heat source layouts} & Reinforcement learning & \rule{0pt}{1.2em}\textbf{\boldmath$5.20\times 10^{-5}$}
 \\ \cline{2-3} 
 & Random Search & \rule{0pt}{1.2em} $6.79\times {{10}^{-5}}$ \\ \cline{2-3} 
 & DARTS & \rule{0pt}{1.2em} $6.14\times {{10}^{-5}}$ \\ \cline{2-3} 
 & ENAS & \rule{0pt}{1.2em}$6.89\times {{10}^{-5}}$ \\ \hline
\multirow{3}{*}{Darcy flow with different input property fields} & Reinforcement learning & \rule{0pt}{1.2em}\textbf{0.0024} \\ \cline{2-3} 
& Random Search & \rule{0pt}{1.2em}0.0046 \\ \cline{2-3} 
 & DARTS & \rule{0pt}{1.2em}0.0041 \\ \cline{2-3} 
 & ENAS & \rule{0pt}{1.2em}0.0045 \\ \hline

\multirow{3}{*}{2D Burgers’ equations with initial conditions} & Reinforcement learning & \rule{0pt}{1.2em}\textbf{0.0005} \\ \cline{2-3} 
& Random Search & \rule{0pt}{1.2em}0.0008 \\ \cline{2-3} 
 & DARTS & \rule{0pt}{1.2em}0.0012 \\ \cline{2-3} 
 & ENAS & \rule{0pt}{1.2em}0.0008 \\ \hline
 
 \multirow{3}{*}{Navier-Stokes equations with varying geometries} & Reinforcement learning & \rule{0pt}{1.2em}\textbf{0.3492} \\ \cline{2-3} 
 & Random Search & \rule{0pt}{1.2em}0.4142 \\ \cline{2-3} 
 & DARTS & \rule{0pt}{1.2em}0.5772 \\ \cline{2-3} 
 & ENAS & \rule{0pt}{1.2em}0.5704 \\ \hline
\end{tabular}
}
\caption{\textbf{Comparison of relative ${L^{2}}$ errors on the validation set for different search strategies across six PDE datasets}. Four search strategies were evaluated: reinforcement learning, random search, Differentiable Architecture Search (DARTS \cite{33}), and Efficient Neural Architecture Search (ENAS \cite{37}). The best results are highlighted in bold.}
\label{search_comp}
\end{table*}

\subsection{Architecture search Space comparison}
To demonstrate the effectiveness of our entire-structured search space, we also evaluate the cell-based search space \cite{31}. The detailed introduction of cell-based search space can be seen in Supplementary Note 8. After searching for 20 network architectures within each space, we select the one with the smallest relative $L^{2}$ error on the validation set as the final metric for comparison. Compared with the cell-based search space baseline, the result in Supplementary Note 8 that on the validation set, the prediction error of the entire-structed search space for parametric PDEs is significantly lower than that of the cell-based search space with the average prediction error is reduced by a factor of 3.9.

\subsection{Search strategies comparison}
We compare the policy-based reinforcement learning search strategy with the one-shot strategy, an efficient NAS method that reduces training time and computational resources by training a `supernet' containing all candidate operations. Unlike the multi-trial strategy, which requires separate training for each candidate structure, the one-shot strategy trains a single supernet, from which the optimal sub-network can be internally obtained. In this paper, we used two one-shot strategy methods: ENAS (Efficient Neural Architecture Search) \cite{37} and DARTS (Differentiable Architecture Search) \cite{33}. The detailed introduction can be seen in Supplementary Note 9. To further validate the efficiency of our search method, we also compared our approach with random search under the same search time constraints. 

\textcolor{black}{Table \ref{search_comp} shows a comparison of these search strategies conducted across all six datasets. Compared to random search, our search method achieves an average reduction by a factor of 1.54. It is also found that the multi-trial search strategy based on reinforcement learning has significantly lower errors on the validation set compared to DARTS and ENAS, with an average reduction by a factor of 1.94 and 1.75 respectively.}
Though it takes more time to search, policy-based reinforcement learning is used as the search strategy due to its superior performance. Visualization and further analysis of the efficiency of our search method can be found in the Discussion section.

\section{Discussion}
{
\color{black}
The core advantage of PICNNs lies in their ability to generalize across a family of PDEs by treating variable physical parameters and boundary/initial conditions as model inputs, thus avoiding costly retraining. However, current PICNN models face significant challenges. We observe that models and loss functions tailored to one PDE dataset often fail to converge on others. For instance, the loss function used by the baseline model PDE-surrogate for training Darcy flow fails to avoid trivial solutions when applied to the heat equation, a phenomenon similar to the one illustrated in Figure \ref{combined_figure} (d). Moreover, PICNNs are currently unable to handle spatiotemporal PDEs effectively. To address these challenges, we proposed Auto-PICNN, which automates the search for loss functions and network architectures tailored to the unique physical characteristics of each PDE. By embedding a ConvLSTM module directly into the PICNN structure, Auto-PICNN extends its applicability to spatiotemporal PDEs. Auto-PICNN can also easily be extended to solve higher-dimensional PDEs, such as spatiotemporal problems in 3D space. This extension only requires replacing 2D convolutions with 3D convolutions. Higher-dimensional PDEs are relatively rare in industrial applications as they are more commonly used in fields such as options pricing and quantum mechanics. These PDEs are typically solved using MLP-based PINNs, which suffer from the inefficiency of retraining from scratch.

In the searched architectures (see the search\_struct.json file in our code repository), we observe that for parametric PDEs without source fields, such as heat equations with different boundary conditions and Navier-Stokes equations with varying geometries, larger convolution kernels (e.g., 5x5 or 7x7) are more commonly employed. This may suggest that larger kernels are better suited for capturing broader spatial dependencies in the absence of source terms. On the other hand, for parametric PDEs with source fields, the architectures tend to utilize a hybrid structure of standard convolutions (Conv) and depth-wise separable convolutions (SepConv), combined with various pooling and upsampling techniques. 
For instance, for PDEs like Darcy flow, our AutoML framework automatically identifies the alternating Conv-SepConv structure as its network architecture. This design likely arises because standard convolutions are effective at capturing fine details, while depth-wise separable convolutions excel at representing larger-scale features. By leveraging this hybrid design and considering Darcy flow's multi-scale dynamics, which range from large-scale permeability variations to localized pressure drops, Auto-PICNN can effectively handle challenging spatial transition areas that baseline PICNNs struggle to solve.

\begin{table}[t!]
\centering
\resizebox{\columnwidth}{!}{
\begin{tabular}{c|c|c|c|c}
\hline
\textbf{Dataset} & \textbf{\begin{tabular}[c]{@{}c@{}}Loss function\\ search time\end{tabular}} & \textbf{\begin{tabular}[c]{@{}c@{}}Network architecture\\ search time\end{tabular}} & \textbf{Epochs} & \textbf{Batch size} \\ \hline\hline
\begin{tabular}[c]{@{}c@{}}Heat equations with different\\ boundary conditions\end{tabular} & 2 GPU hours & 3 GPU hours & 1000 & 1 \\ \hline
\begin{tabular}[c]{@{}c@{}}Poisson equations with varying\\ source fields\end{tabular} & 12 GPU hours & 1 GPU day & 10000 & 32 \\ \hline
\begin{tabular}[c]{@{}c@{}}Poisson equations with different\\ heat source layouts\end{tabular} & 6 GPU hours & 1.5 GPU days & 30 & 1 \\ \hline
\begin{tabular}[c]{@{}c@{}}Darcy flow with different\\ input property fields\end{tabular} & 10 GPU hours & 1 GPU day & 300 & 32 \\ \hline
\begin{tabular}[c]{@{}c@{}}2D Burgers’ equation\\ with initial conditions\end{tabular} & 12 GPU hours & 1.5 GPU day & 5000 & 100 \\ \hline
\begin{tabular}[c]{@{}c@{}}Navier-Stokes equations with\\ varying geometries\end{tabular} & 12 GPU hours & 1 GPU day & 20000 & 1 \\ \hline
\end{tabular}
}
\caption{\textbf{The search time for loss function and network architecture, training epochs, and batch size for Auto-PICNN across different PDE datasets.
}}
\label{searchcost}
\end{table}

To comprehensively evaluate the performance of Auto-PICNN, we compared our approach not only with traditional PICNN baselines but also with state-of-the-art neural operator methods that have emerged as powerful alternatives for solving parametric PDEs. While Auto-PICNN shares similarities with neural operator methods like DeepONet and FNO in being designed for surrogate modeling of parametric PDEs, a fundamental difference lies in their training paradigms. DeepONet and FNO are data-driven approaches that require generating large amounts of ground truth solutions, consuming substantial computational time and resources. In contrast, PICNNs are PDE-driven, training directly on governing equations without requiring any ground truth solutions in the training stage. Nevertheless, to provide a comprehensive evaluation against these established methods, we compared Auto-PICNN with DeepONet and FNO using identical training, validation, and test sets across all six datasets. The results demonstrate that Auto-PICNN significantly outperforms both methods even without access to ground truth solutions during training. Specifically, Auto-PICNN achieves an average error reduction of 14.90-fold compared to FNO and 25.37-fold compared to DeepONet across the six diverse PDE systems, as detailed in Supplementary Note 10.

Another key advantage of Auto-PICNN over DeepONet and FNO is its minimal hyperparameter requirements due to automated search. 
Since Auto-PICNN automatically selects both loss functions and network architectures including its specific convolution types, our framework requires manual setting of very few hyperparameters. Taking the learning rate as an example of such hyperparameters, we used the heat equation with different boundary conditions to test the framework's sensitivity to learning rate variations. As shown in Supplementary Note 11, the results demonstrate that regardless of different learning rate settings, the prediction errors remain consistently low, highlighting the framework's robustness and practical applicability.

The practicality of our framework is further demonstrated by its high search efficiency, which starkly contrasts with the challenges of manual design. These challenges primarily stem from domain scientists often lacking expertise in both loss function formulation and network architecture search space construction, necessitating extensive manual experimentation. Even when provided with well-designed search spaces, they may not know how to effectively search for optimal architectures, leading to prolonged trial-and-error periods that consume substantial time and computational resources. These time costs are non-negligible and may extend to weeks or even months without achieving satisfactory results. In contrast, our AutoML approach addresses these challenges by constructing appropriate search spaces and employing efficient search strategies rather than exhaustive enumeration. The search time for loss functions and network architectures, along with the batch sizes that are identical to those used in baseline models for each PDE dataset can be seen in Table \ref{searchcost}.
As can be seen, our search times are controlled within approximately two days for even the most complex cases, which is highly efficient in practice, especially considering that manual design typically requires much longer periods with inferior results.

\begin{figure}[t]
    \centering
    \includegraphics[width=0.99\linewidth]{Figure/combined_rl_curve.jpg}
    \caption{\textbf{Reinforcement learning-based network architecture search process for the heat equation.} Subfigures (a), (b), (c), (d), and (e) illustrate the search process for gradually increasing sizes of the training dataset, with training sample sizes of 2, 3, 4, 5, and 6 samples, respectively. Each subfigure shows how the reward (reciprocal of relative $L^{2}$ error) evolves during the reinforcement learning process as the controller learns to generate better network architectures.}
    \label{conver_re}
\end{figure}

\begin{table}[t]
\centering
\resizebox{\columnwidth}{!}{
\begin{tabular}{c|c|c|c}
\hline
\textbf{Dataset} & \textbf{Model} & \textbf{Retrain (s)} & \textbf{Inference (ms/batch)} \\ \hline\hline
\multirow{2}{*}{Heat equations with different boundary conditions} & PhyGeoNet & 29.10 & 1.22 \\ \cline{2-4}  
 & Auto-PICNN & \textbf{28.9} & \textbf{0.82} \\ \hline
\multirow{2}{*}{Poisson equations with varying source fields} & PhyGeoNet & 20900.6 & 2.66 \\ \cline{2-4}  
 & Auto-PICNN & \textbf{1875.4} & \textbf{2.30} \\ \hline
\multirow{2}{*}{Poisson equations with different heat source layouts} & PI-UNet & 3672 & 4.43 \\ \cline{2-4}  
 & Auto-PICNN & \textbf{3504} & \textbf{3.36} \\ \hline
\multirow{2}{*}{Darcy flow with different input property fields}
 & PDE-surrogate  & 1667.4 & 5.23 \\ \cline{2-4} 
 & Auto-PICNN & \textbf{1316.4} & \textbf{3.82} \\ \hline
\multirow{2}{*}{2D Burgers' equations with initial conditions}
 & PhyCRNet  & 13200.1 & \textbf{355} \\ \cline{2-4} 
 & Auto-PICNN & \textbf{7650.5} & 375 \\ \hline
\multirow{2}{*}{Navier-Stokes equations with varying geometries} & PhyGeoNets & 5416.8 & 0.42 \\ \cline{2-4} 
 & Auto-PICNN & \textbf{1275.8} & \textbf{0.15} \\ \hline
\end{tabular}
}
\caption{\textbf{Comparison of retraining and inference time between baseline models and Auto-PICNN across 6 PDE datasets.} Retraining times are reported in seconds (s), and inference times are shown in milliseconds per batch (ms/batch). The best results are highlighted in bold.}
\label{retrain}
\end{table}

The efficiency and effectiveness of our two-stage search strategy are also well-supported by both theoretical guarantees and empirical evidence. The use of Bayesian optimization with Gaussian process for the loss function search is backed by strong theoretical guarantees for finding near-optimal solutions by Srinivas et al. \cite{srinivas2009gaussian}, ensuring that the identified loss function is both robust and effective. For the network architecture search, in line with existing research in RL-based neural architecture search that relies on empirical validation \cite{36,jaafra2019reinforcement,salehin2024automl}, we also demonstrate its effectiveness empirically. This is visually presented in Figure \ref{conver_re}, which uses the heat equation with different boundary temperatures as a case study. 
The figure illustrates the search process as the training set size progressively increases from two samples to six. It is evident that across all dataset sizes, the reinforcement learning strategy consistently guides the search towards higher rewards (the reciprocal of prediction error which can be seen in Equation \ref{eq:reward}), and ultimately converges on a solution with minimal prediction error. Notably, we empirically observe that when excluding the increased training time that naturally comes with larger datasets, the search time itself is inversely proportional to the size of the dataset. This suggests that the larger the size of the PDE dataset, the fewer search iterations are needed to identify the optimal architecture. Beyond the efficiency of the search process itself, the architectures discovered by Auto-PICNN also demonstrate superior computational performance. Compared to baseline models, our searched architectures achieve both faster convergence during retraining and reduced inference times, as shown in Table \ref{retrain}. This advantage is particularly pronounced for retraining, where our automatically designed architectures achieve significantly better performance while requiring substantially less training time.

To further validate the robustness of our approach across different scientific domains, we extended our evaluation to include reaction-diffusion systems. Specifically, we tested Auto-PICNN on the 2D Gray-Scott equations, a particularly stiff PDE system that exhibits complex spatiotemporal patterns arising from the interplay between chemical reactions and diffusion processes. The dataset was sourced entirely from PeRCNN \cite{rao2023encoding}. We selected the first 200 timesteps, covering 100 seconds, as the training set, with the last 50 timesteps of this set used for validation. The subsequent timesteps from 200 to 300 served as the test set. Auto-PICNN achieved a relative $L^{2}$ error of 0.062 on the test set. Compared to the baseline PeRCNN model \cite{rao2023encoding}, which achieved a relative $L^{2}$ error of 0.145, Auto-PICNN delivered a 57.24\% error reduction. Furthermore, even though Auto-PICNN does not utilize ground truth information during the training phase, it still outperformed these neural operator methods, achieving error reductions of 66.67\% and 4.62\% relative to DeepONet and FNO, respectively.
These results confirm that our automated framework maintains its effectiveness even for challenging stiff systems that span beyond physics applications into chemical and biological domains.

\textcolor{black}{Ultimately, to better contextualize our Auto-PICNN framework within the broader landscape of physics-informed machine learning, we present a detailed comparison with existing NAS applications to PINNs \cite{24, 25,40}. Among these prior works, NAS-PINN \cite{40} represents the most recent and advanced approach, employing techniques like DARTS for MLP architecture optimization. To articulate the specific advancements of our framework over these methods, we systematically analyzed the key differences and our unique technical innovations in Supplementary Note 16. This analysis demonstrates that our approach provides advancements for the physics-informed machine learning field, extending beyond a direct application of existing NAS techniques. Building on this, to further demonstrate the broader applicability of our innovations beyond CNN-based models, we also transferred our operator-infused loss function search space to traditional MLP-based PINNs. Specifically, we compared our approach against both vanilla PINNs and NAS-PINN in the traditional PINN context. Our experiments show substantial performance improvements, achieving an average 71.27\% error reduction compared to vanilla PINNs and 53.71\% error reduction compared to NAS-PINN across three diverse datasets. This result also confirms the powerful, generalizable nature of our loss function search paradigm, with the detailed experiment and analysis provided in Supplementary Note 17.}

While Auto-PICNN demonstrates promising results in solving various PDEs, we acknowledge certain limitations in our current approach. Specifically, decoupling the search for loss functions and network architectures may lead to locally optimal solutions. In the field of AutoML, there are currently very few studies dedicated to combining neural architecture search and hyperparameter optimization. In future work, the loss function and network architecture of PICNNs may be jointly searched to achieve globally optimal solutions, further enhancing the accuracy and robustness of the framework.
}

\section{Methods}
\subsection{Physics-Informed Convolutional Neural Networks (PICNNs)}

Consider a parametric PDE in the spatial domain:
\begin{equation}
\mathcal{F}\left[ u\left( \textbf{x} \right),\mu  \right]=f\left( \textbf{x} \right),\quad \textbf{x}\in \Omega 
\end{equation}\label{Equ_1}
\begin{equation}
\mathcal{B}\left[ u\left( \textbf{x} \right),\mu  \right]=g\left( \textbf{x} \right),\quad \textbf{x}\in \partial \Omega
\end{equation}\label{Equ_2}
where $u\left( \textbf{x} \right)$ is the solution of the parametric PDE, $\mathcal{F}$ is the differential operator of the PDE, $\mathcal{B}$ is the boundary condition operator, $\Omega$ is the domain of the equation, and $\partial \Omega$ is the boundary of the domain. The parameter $\mu$ can represent either the parameter-dependent fields (see detailed construction method in \cite{12}) or the source fields $\mu(\textbf{x} )$ of the PDE. In PICNNs, the CNN takes fields parameterized by $\mu$ as input, which are sampled at discrete points (known as collocation or residual points) within the spatial domain. After passing through the CNN, the output represents the corresponding solution values of the PDE at these points. The construction of the loss function is as follows:
\begin{equation}
    L\left( \theta  \right)={{\lambda }_{{\mathrm{r}}}}{{L}_{{\mathrm{r}}}}\left( \theta  \right)+{{\lambda }_{\mathrm{b}}}{{L}_{{\mathrm{b}}}}\left( \theta  \right)
\end{equation}
\begin{equation}
    L_{\mathrm{r}}(\theta) = \frac{1}{N_{\mathrm{r}}} \sum_{n=1}^{N_{\mathrm{r}}} \eta _{\mathrm{r}}^n \left| \mathcal{F}\left[u_\theta \left(\mathbf{x}_{\mathrm{r}}^n \right)\right] - f\left(\mathbf{x}_{\mathrm{r}}^n\right) \right|^2
\end{equation}
\begin{equation}
    L_{\mathrm{b}}(\theta) = \frac{1}{N_{\mathrm{b}}} \sum_{n=1}^{N_{\mathrm{b}}} \eta _{\mathrm{b}}^n \left| \mathcal{B}\left[u_\theta \left(\mathbf{x}_{\mathrm{b}}^n \right)\right] - g\left(\mathbf{x}_{\mathrm{b}}^n\right) \right|^2
\end{equation}
where $\left\{\mathbf{x}_{\mathrm{r}}^{n} \right\}_{n=1}^{{{N}_{{\mathrm{r}}}}}$ are the residual points sampled within the domain $\Omega$ and $\left\{\mathbf{x}_{\mathrm{b}}^{n} \right\}_{n=1}^{{{N}_{{\mathrm{b}}}}}$ are the boundary points with ${N}_{\mathrm{r}}$ and ${N}_{\mathrm{b}}$ being the number of residual points and boundary points. $\left\{ \eta _{\mathrm{r}}^n \right\}_{n=1}^{{{N}_{r}}}$ are the coefficients of the residual loss at each point in the domain $\Omega$, while $ \left\{ \eta _{\mathrm{b}}^n \right\}_{n=1}^{{{N}_{b}}}$ are the coefficients of the boundary loss at each point on the boundary $\partial \Omega$. Typically, in standard PICNNs, the values of both $\left\{ \eta _{\mathrm{r}}^n \right\}_{n=1}^{{{N}_{\mathrm{r}}}}$ and $\left\{ \eta _{\mathrm{b}}^n \right\}_{n=1}^{{{N}_{\mathrm{b}}}}$ are set to 1. ${{\lambda }_{\mathrm{r}}}$, ${{\lambda }_{\mathrm{b}}}$ are coefficients of the PDE residual term ${{L}_{\mathrm{r}}}$, boundary condition loss term ${{L}_{\mathrm{b}}}$. ${{u}_{\theta }}\left( x\right)$ represents the CNN used to approximate the solution of the PDEs with $\theta$ being the parameters of CNNs. In PICNNs, to calculate the spatial derivatives of the neural network, instead of using the automatic differentiation in MLP-based PINNs, PICNNs use numerical methods (see the following subsection `Calculation of spatial derivatives' for details) to compute the partial derivatives of the $u(\mathbf{x})$.

To extend PICNNs to solve spatiotemporal PDEs, the formulation incorporates the temporal domain $t \in [0, T]$, and the parametric PDE is represented as:

\begin{equation}
\mathcal{F}\left[u\left(\mathbf{x}, t\right), \mu \right] = f\left(\mathbf{x}, t\right), \quad \mathbf{x} \in \Omega, \; t \in [0, T],
\end{equation}
\begin{equation}
\mathcal{B}\left[u\left(\mathbf{x}, t\right), \mu \right] = g\left(\mathbf{x}, t\right), \quad \mathbf{x} \in \partial \Omega, \; t \in [0, T],
\end{equation}
\begin{equation}
\mathcal{I}\left[u\left(\mathbf{x}, 0\right)\right] = h\left(\mathbf{x}\right), \quad \mathbf{x} \in \Omega,
\end{equation}
where $\mathcal{I}$ specifies the initial condition of the PDE. To model spatiotemporal PDEs, a ConvLSTM module is embedded within the intermediate layers of the CNN to capture temporal dynamics effectively. Let $\mathbf{Y}_t$ denote the output of the intermediate layer of the CNN at time step $t$. The ConvLSTM updates the cell state $\mathbf{C}_t$ and hidden state $\mathbf{h}_t$ as follows:

\begin{equation}
\mathbf{i}_t = \sigma\left(\mathbf{W}_{\mathrm{i}} \ast \left[\mathbf{Y}_t, \mathbf{h}_{t-1}\right] + \mathbf{b}_\mathrm{i}\right),
\end{equation}
\begin{equation}
\mathbf{f}_t = \sigma\left(\mathbf{W}_\mathrm{f} \ast \left[\mathbf{Y}_t, \mathbf{h}_{t-1}\right] + \mathbf{b}_\mathrm{f}\right),
\end{equation}
\begin{equation}
\tilde{\mathbf{C}}_t = \mathrm{tanh}\left(\mathbf{W}_\mathrm{c} \ast \left[\mathbf{Y}_t, \mathbf{h}_{t-1}\right] + \mathbf{b}_\mathrm{c}\right),
\end{equation}
\begin{equation}
\mathbf{C}_t = \mathbf{f}_t \odot \mathbf{C}_{t-1} + \mathbf{i}_t \odot \tilde{\mathbf{C}}_t,
\end{equation}
\begin{equation}
\mathbf{o}_t = \sigma\left(\mathbf{W}_\mathrm{o} \ast \left[\mathbf{Y}_t, \mathbf{h}_{t-1}\right] + \mathbf{b}_\mathrm{o}\right),
\end{equation}
\begin{equation}
\mathbf{h}_t = \mathbf{o}_t \odot \mathrm{tanh}\left(\mathbf{C}_t\right),
\end{equation}
where $\mathbf{i}_t$, $\mathbf{f}_t$, and $\mathbf{o}_t$ represent the input, forget, and output gates, respectively. The operators $\ast$ and $\odot$ denote convolution and element-wise multiplication, respectively. The weights $\mathbf{W}_\mathrm{i}, \mathbf{W}_\mathrm{f}, \mathbf{W}_\mathrm{c}, \mathbf{W}_\mathrm{o}$ and bias terms $\mathbf{b}_\mathrm{i}, \mathbf{b}_\mathrm{f}, \mathbf{b}_\mathrm{c}, \mathbf{b}_\mathrm{o}$ are shared across time steps.

For such problems, the loss function incorporates an additional term for the initial condition:

\begin{equation}
L(\theta) = \lambda_{\mathrm{r}} L_{\mathrm{r}}(\theta) + \lambda_{\mathrm{b}} L_{\mathrm{b}}(\theta) + \lambda_\mathrm{i} L_\mathrm{i}(\theta),
\end{equation}
\begin{equation}
L_\mathrm{i}(\theta) = \frac{1}{N_\mathrm{i}} \sum_{n=1}^{N_\mathrm{i}} \eta_\mathrm{i}^n \left|\mathcal{I}\left[u_\theta\left(\mathbf{x}_\mathrm{i}^n, 0\right)\right] - h\left(\mathbf{x}_\mathrm{i}^n\right)\right|^2,
\end{equation}
where $N_\mathrm{i}$ is the number of initial condition points, and $\eta_\mathrm{i}^n$ are their respective coefficients. The input of the CNN in this case is the initial condition $h\left(\mathbf{x}\right)$. 

This framework ensures that PICNNs are not only capable of solving steady-state PDEs but also spatiotemporal PDEs by leveraging both spatial and temporal modeling.

\subsection{Loss function construction}
In PICNNs, the vanilla MSE loss function (see Equation (3) or Equation (15)) is commonly used but the construction of the proper loss function may vary for different PDEs. To reduce prediction error, it is necessary to find a suitable loss function for a specific PDE dataset. The construction of the loss function may be influenced by the following four factors.

\subsubsection{Boundary conditions}
There are two types of constraints on the boundary conditions used in PINNs and PICNNs. One is the soft constraint shown in Equations (4), (5) and (8). The other is the hard constraint that modifies the network outputs such that the boundary conditions can be strictly imposed into the network architecture during training. \textcolor{black}{While existing PICNNs typically choose one fixed approach (either soft or hard constraints), the operator-infused loss function search space of Auto-PICNN can automatically search between these two constraint types to determine the optimal boundary constraint for each specific PDE.} Furthermore, our search space, beyond these two constraint types, as detailed in the subsequent subsection `operator-infused loss function search space', also encompasses an approach that combines both soft and hard constraints. It has the penalty term of the boundary condition added to the loss function while ensuring that the hard constraints are strictly imposed before the differentiation operations of the PDEs.

\subsubsection{Calculation of spatial derivatives}
For PDEs in the spatial domain, Sobel filters and finite difference methods can be used to calculate spatial derivatives of different orders. The core idea is to use predefined, untrainable kernels to convolve the model output matrix. This convolution process enables the computation of horizontal and vertical gradients of the PDEs at each spatial position. In this paper, four kernels (3x3 Sobel filter, 5x5 Sobel filter, $2^{nd}$ order accuracy central difference, $4^{th}$ order accuracy central difference) are employed to calculate spatial differentiation with different orders. A detailed description of the four convolution kernels can be found in Supplementary Note 12.

\subsubsection{Regression loss function}
The $L^{2}$ physics-informed loss known as the mean squared error (MSE) is widely used in both PINNs and PICNNs. However, \cite{41} pointed out that $L^{2}$ physics-informed loss may not be a suitable choice for some PDEs. Instead, they chose $L^{{\infty}}$ as the form of the regression loss function. Besides, the $L^{1}$ physics-informed loss known as the mean absolute error (MAE) has also been used \cite{13} in training. Therefore, our search space also considers this factor.

\subsubsection{Weighting strategies for different residual points}
More and more studies have made advancements in enhancing the components and weights balancing within the vanilla loss function of PINNs based on previous work \cite{30,19,13}. However, these methods are mainly used in PINNs instead of PICNNs. Inspired by them, we design several operators in the search space and make sure that the ideas of these methods can be comprised in our search space. Detailed illustrations of how these weighting strategies can be combined in PICNNs and their corresponding operators in our search spaces can be seen in Supplementary Note 13. 
\begin{figure}[h!]
    \centering
    \includegraphics[width=0.85\linewidth,clip,trim=0.1cm 0.5cm 0.1cm 0.1cm]{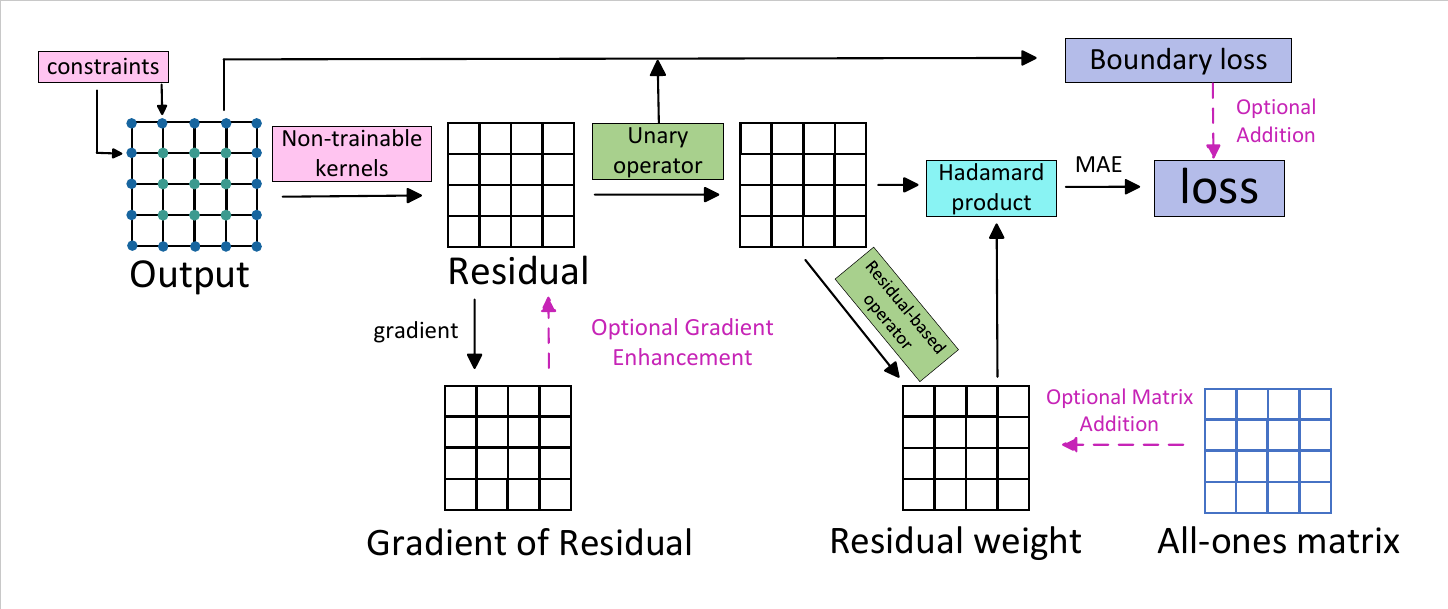}

     \caption{\textbf{The operator-infused loss function search space.} The search space consists of seven key components, each offering multiple operational choices. Starting from the network output, the pipeline includes: (1) constraint component determining whether hard constraints are applied to enforce boundary conditions; (2) which non-trainable convolution kernel to use for computing spatial derivatives and residuals; (3) which unary operator (absolute value, squaring, or identity) to apply to the residual matrix; (4) whether to include gradient enhancement of the residual; (5) which residual-based operator (top-N thresholding, normalization, unitization, or pointwise loss gradient ascent) to apply for generating weight matrices; (6) whether to augment weights with an all-ones matrix; and (7) whether to integrate boundary loss. The final loss function is constructed by selecting one operation from each component.}
    \label{loss search}
\end{figure}

\subsubsection{Operator-infused loss function search space}

Figure \ref{loss search} shows the overall construction of our operator-infused loss function search space for PICNNs. This search space innovatively combines a variety of components and operators for the computation of loss functions. The total loss function in our operator-infused loss function search space is defined as:
\begin{align}
\mathcal{L}_\mathrm{total} = &\ \mathrm{MAE}\left( \mathbf{W}_\mathrm{final} \odot U\left( \mathbf{K} \ast \mathbf{O}_\mathrm{constrained} \right) \right) \nonumber  + \lambda_{\mathrm{b}} \sum_{i \in \partial \Omega} U\left( \mathbf{O}_i - g(\mathbf{x}_i) \right) \nonumber \\
& + \lambda_\mathrm{g} \mathrm{MAE}\left( \mathbf{W}_\mathrm{gradient} \odot U\left( \|\nabla \mathbf{R}\| \right) \right),
\end{align}
where:
\begin{itemize}
    \item $\mathcal{L}_\mathrm{total}$: Total loss for training the PICNN model.
    \item $\mathbf{O}$: The output of the neural network, representing the predicted solution of PDEs.
    \item $\mathbf{O}_\mathrm{constrained}$: The modified or unmodified output of the network, depending on whether a hard constraint is applied:
    \begin{equation}
    \mathbf{O}_\mathrm{constrained} =
    \begin{cases}
        f_\mathrm{hard}(\mathbf{O}), & \text{if hard constraint is applied}, \\
        \mathbf{O}, & \text{otherwise}.
    \end{cases}
\end{equation}
    Here, $f_\mathrm{hard}(\cdot)$ strictly imposes boundary conditions (e.g., Dirichlet or Neumann constraints) on the network output by directly modifying $\mathbf{O}$.
    \item $\mathbf{K}$: Non-trainable convolution kernel (e.g., Sobel filter or finite difference kernel).
    \item $\mathbf{R} = \mathbf{K} \ast \mathbf{O}_\mathrm{constrained}$: Residual matrix, computed by convolving $\mathbf{O}_\mathrm{constrained}$ with $\mathbf{K}$, where $\ast$ denotes the convolution operation.
    \item $U(\cdot)$: Unary operator applied to transform elements of a matrix.
    \item $\mathbf{W}_\mathrm{final} = \mathbf{W}_\mathrm{residual} + \alpha \mathbf{1}$: Final weight matrix, combining residual weights derived from residual-based operator with optional augmentation by an all-ones matrix $\mathbf{1}$ with weight $\alpha$.
    \item $\lambda_{\mathrm{b}}$: Weight for the boundary loss term.
    \item $g(\mathbf{x}_i)$: True boundary condition at position $\mathbf{x}_i$, provided by the problem setup.
    \item $\lambda_\mathrm{g}$: Weight for the gradient term.
    \item $\|\nabla \mathbf{R}\|$: Magnitude of the residual gradient vector:
    \begin{equation}
    \|\nabla \mathbf{R}\| = \sqrt{\left(\frac{\partial \mathbf{R}}{\partial x}\right)^2 + \left(\frac{\partial \mathbf{R}}{\partial y}\right)^2}.
    \end{equation}
    \item $\mathbf{W}_\mathrm{gradient}$: Weight matrix for the gradient term derived from residual-based operator.
    \item $\odot$: Hadamard product (element-wise multiplication).
    \item $\mathrm{MAE}(\cdot)$: Mean Absolute Error applied to element-wise weighted residual matrices.
\end{itemize}

\textcolor{black}{The formulation of our loss function in Equation (17) can be visualized through the construction process of the operator-infused loss function search space. As shown in Figure \ref{loss search}, each key component within this loss function search space offers several distinct operational choices and allows for flexibility in tailoring the loss to specific PDE characteristics. The PICNN for each PDE will have its own better-suited selection among these options, as different PDEs may favor distinct loss formulations to better capture their unique physical dynamics. Through our automated search process, for each PDE, we select the optimal operation from the available options within each component and combine them to form a tailored loss function. Each component of the loss function, along with its set of available operations, is described in detail below:}
\begin{enumerate}
\item \textbf{Constraint ($\mathbf{O}_\mathrm{constrained}$):}  
   \textcolor{black}{Before the model's output is used to compute the PDE residuals via convolutional differentiation, the constraint component of our search space first determines how boundary conditions are handled. This component offers two mutually exclusive options: either a hard constraint is imposed on the model output, or no hard constraint is applied, which defaults to a soft constraint. This choice directly corresponds to the two primary approaches for handling boundary conditions currently used in the PICNN literature, as we discuss in the subsection `Boundary conditions'.}

\item \textbf{Non-Trainable Kernels:}
    \textcolor{black}{We offer several options for non-trainable convolution kernels, which are discussed in subsection `Calculation of spatial derivatives' and Supplementary Note 12, allowing the search to select the most suitable one for performing spatial differentiation of various orders.} This process yields the residual matrix, where each element ${r}_{i,j}$ represents the PDE residual loss at its corresponding position $(i, j)$ in the domain $\Omega$.
    
\item \textbf{Unary operators on residual:} Unary operators $U(\cdot)$ transform the residual matrix $R$:
\begin{equation}
R_\mathrm{u} = U(R),
\end{equation}
\textcolor{black}{where $U(\cdot)$ includes three options: the absolute value ($|\cdot|$), squaring ($(\cdot)^2$), or the identity function ($\cdot$). This allows the emphasis on the residuals to be customized for different PDEs.}

\item \textbf{Optional gradient enhancement of residual:} \textcolor{black}{This component of the search space determines whether to enhance the residual with its gradient in the loss function, offering two distinct options: include the gradient term or exclude it ($\lambda_\mathrm{g} = 0$). This idea is inspired by GPINNs \cite{30}. During the training process for the PICNN corresponding to a specific PDE, this allows judgment on whether adding a loss term based on the residual's gradient is needed, which forces the PDE residual to be not only small but also flat, enabling the search to determine if such a smoother residual constraint is beneficial for the particular PDE being solved.}

\item \textbf{Residual-based operators for weighting:} 
Residual-based operators generate the weight matrix $W_\text{residual}$ from $R_u$. Possible operations include:
\begin{enumerate}[label=(\roman*)]
    \item \textbf{Top-$N$ Thresholding:}
    \begin{equation}
    (W_\mathrm{residual})_{i,j} =
    \begin{cases} 
        1, & \text{if } (R_u)_{i,j} \in \text{Top-}N, \\
        0, & \text{otherwise}.
    \end{cases}
    \end{equation} 
\textcolor{black}{This design enables the search to determine whether it is beneficial to concentrate training only on points with larger residual errors while ignoring those with smaller errors.}

    \item \textbf{Normalization:}
    \begin{equation}
    (W_\mathrm{residual})_{i,j} = 
    \frac{(R_\mathrm{u})_{i,j} - \min(R_\mathrm{u})}{\max(R_\mathrm{u}) - \min(R_\mathrm{u})}.
    \end{equation}
    \textcolor{black}{This design enables the search to determine whether assigning higher weights to points with larger residual errors and lower weights to those with smaller errors is beneficial during training.}

    \item \textbf{Unitization:}
    Sets all elements of $W_\mathrm{residual}$ to 1:
    \begin{equation}
    W_\mathrm{residual} = 1.
    \end{equation}
    \textcolor{black}{This is the vanilla MSE loss approach for handling residual loss, assigning equal weight to every point regardless of its residual value.}

    \item \textbf{Pointwise Loss gradient ascent:}  
This operation, inspired by \cite{39}, calculates the derivative of the loss function with respect to the current weight of each residual point. Each residual point starts with an initial weight of 1, and the weights are updated iteratively using gradient ascent. The update rule for the residual-based weight matrix $\mathbf{W}_\mathrm{residual}$ can be written as:
\begin{equation}
\mathbf{W}_\mathrm{residual}^{k+1} = \mathbf{W}_\mathrm{residual}^k + \rho_k^r \nabla_{\mathbf{W}_\mathrm{residual}} L(\mathbf{w}^k, \mathbf{W}_\mathrm{residual}^k),
\end{equation}
where:
\begin{itemize}
    \item $\mathbf{w}^k$: The weights of the neural network at iteration $k$.
    \item $\mathbf{W}_\mathrm{residual}^k$: The residual-based weight matrix at iteration $k$, initialized to 1.
    \item $\rho_k^r$: Learning rate for updating the residual-based weight matrix.
    \item $L(\mathbf{w}^k, \mathbf{W}_\mathrm{residual}^k)$: The loss function evaluated with the current neural network weights $\mathbf{w}^k$ and residual weights $\mathbf{W}_\mathrm{residual}^k$.
    \item $\nabla_{\mathbf{W}_\mathrm{residual}}$: Gradient of the loss function with respect to the residual-based weight matrix $\mathbf{W}_\mathrm{residual}$.
\end{itemize}
    
\end{enumerate}

\item \textbf{Optional matrix addition:} 
The weight matrix may be optionally augmented with an all-ones matrix. When $\alpha$ is set to 0 in $\mathbf{W}_\mathrm{final} = \mathbf{W}_\mathrm{residual} + \alpha \mathbf{1}$, it indicates that there is no augmentation applied

\item \textbf {Loss boundary addition:} The integration of boundary loss into the final loss computation is optional. When $\lambda_{\mathrm{b}}$ in Equation (17) is set to 0, it indicates that no boundary loss is included in the loss function.
\end{enumerate}

This operator-infused loss function search space not only incorporates concepts from existing strategies used in PINNs (see Supplementary Note 13) but also explores a wide range of innovative strategies in loss function formulation.

\subsection{Search space for the network architecture}
\begin{figure}[t!]
    \centering
    
    \includegraphics[width=0.97\linewidth]{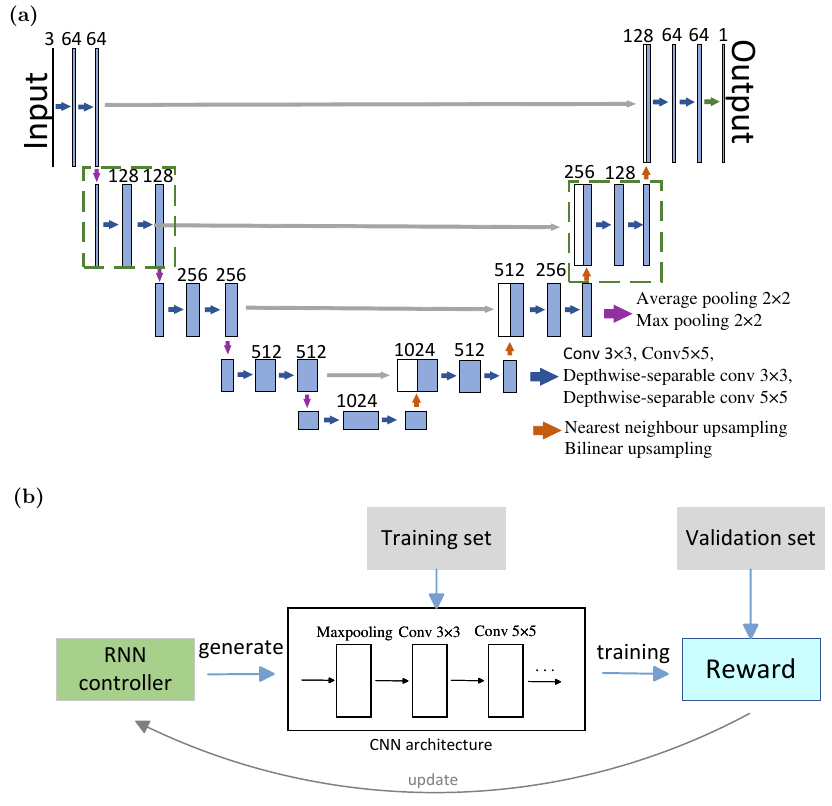}
    \caption{\textbf{Network architecture search space and search strategy.} (a) shows our entire-structured search space constructed based on UNet architecture, where each layer can select from multiple candidate operations. (b) illustrates the policy-based reinforcement learning process, where a recurrent neural network controller generates candidate architectures and updates its parameters based on the resulting reward, which is the reciprocal of relative $L^{2}$ error on the validation set.}
    \label{fig:combined}
\end{figure}

Following \cite{13} and \cite{rahman2022u}, for models solving PDEs with varying source fields, we choose the classic image segmentation model UNet \cite{27} as the backbone model, 
and use the entire-structured search space \cite{31}. In this approach, to find a network architecture with the best performance, we traverse the whole search space of architecture, considering the operations at each layer. Figure \ref{fig:combined} (a) shows the search space based on the entire structure of UNet.

For the upsampling operation, \cite{11} pointed out that using transposed convolution for upsampling would lead to worse prediction results. Based on this, this paper defines the following candidate operations for the convolutional layers, upsampling, and downsampling of UNet:
\begin{enumerate}
  \item Convolutional layer: Standard convolutions with kernel size of 3$\times$ 3 and 5$\times$5, depth-wise separable convolutions with kernel size of 3$\times$3 and 5$\times$5.
  \item Upsampling: Bilinear interpolation, Nearest-neighbor interpolation.
  \item Downsampling: Max pooling, average pooling.
\end{enumerate}
Besides, Group Normalization and GeLU activation functions are used as normalization and activation functions. Additionally, for PDEs with different boundary conditions or geometries that have no input source fields, a slightly modified search space architecture is employed, as detailed in Supplementary  Note 14.

  
  
  

\subsection{Two-stage search strategy for PICNNs}
\cite{1} shows that improper selection of the loss function can have a significant impact on prediction performance. The optimal architecture searched under an inappropriate loss function cannot effectively improve prediction results. Based on this, we adopt a two-stage search strategy, dividing the search process into two stages: 1) loss function search stage and 2) network architecture search stage. The implementation details of these models are provided in the Supplementary
Note 10.
\subsubsection{Loss Function Search Stage}
All components as well as operations in our search space can be regarded as hyperparameters affecting the loss functions thus hyperparameter optimization can be conducted. For parametric PDEs, the default structure for loss function search stage utilizes the standard 3x3 convolution and 2x2 max pooling operations from the vanilla UNet model. In this paper, the Bayesian optimization based on Gaussian processes \cite{34} and the median stopping strategy \cite{35} are adopted for hyperparameter optimization of the search space. The fundamental principle behind Bayesian optimization is to estimate an unknown function by utilizing a prior distribution and then iteratively select sampling points based on an acquisition function defined by the posterior distribution. The median stopping strategy can be used to terminate their training prematurely. The detailed introduction of these two methods can be seen in Supplementary Note 15.

\subsubsection{Physics informed architecture search}
After determining the form of the loss function, we searched for the network architecture. We adopt the multi-trial strategy and use policy-based reinforcement learning \cite{36}. The architecture search problem is framed as a reinforcement learning problem, where the task is to train a controller that generates model architectures for PICNNs. The controller is implemented using the LSTM, which progressively generates descriptions of neural network components from the search space. Once the controller generates the entire neural network structure, the reciprocal of the relative ${L^{2}}$ error $e$ on the validation set of the PICNNs will be considered as the reward in reinforcement learning, i.e.:
\begin{equation}
    e=\frac{\sqrt{\mathop{\sum }_{i=1}^{N}{{\left| {{u}_{\theta }}\left( {{\mathbf{x}}_{i}} \right)-u\left( {{\mathbf{x}}_{i}} \right) \right|}^{2}}}}{\sqrt{\mathop{\sum }_{i=1}^{N}{{\left| u\left( {{\mathbf{x}}_{i}} \right) \right|}^{2}}}}
\end{equation}
\begin{equation}
    \mathrm{reward}=\frac{1}{e}
    \label{eq:reward}
\end{equation}
where ${{u}_{\theta }}\left( {{\mathbf{x}}_{i}} \right)$ and $u\left( {{\mathbf{x}}_{i}} \right)$ are the prediction of PICNNs and the high-fidelity solution to the PDEs, respectively. The reciprocal of $e$ is chosen as the reward to transform the error minimization into a maximization problem, which aligns with the common reinforcement learning setting.

As shown in Figure \ref{fig:combined} (b), by using the rewards of the generated PICNN model to update the policy gradient, the policy-based reinforcement learning can learn to optimize the controller parameters $\theta$ so that the controller can find the optimal CNN architecture.

Specifically, the policy-based reinforcement learning gives the approximation of the policy gradient of $\theta$:
\begin{equation}
    {{\nabla }_{\theta }}J\left( \theta  \right)=\mathop{\sum }_{t=1}^{T}{{\nabla }_{\theta }}\log \pi \left( {{a}_{t+1}}\left| {{a}_{t}};\theta  \right. \right)\times \mathrm{reward}
\end{equation}
where $T$ is the number of components the controller has to generate, ${{a}_{t}}$ is the component of the neural network generated by the controller from the candidate operations at step $t$, and $J\left( \theta  \right)$ represents the expected cumulative rewards. Then, the policy parameters are updated using the stochastic gradient ascent method:
\begin{equation}
    \theta \leftarrow \theta +\alpha{{\nabla }_{\theta }}J\left( \theta  \right)
\end{equation}
where $\alpha$ is the learning rate. The controller continuously generates new CNN architectures until convergence. Finally, the model with the highest reward is selected as the final CNN model.


\section{Data availability}
All the used datasets in this study can be found on GitHub: \url{https://github.com/wanyunzh/AutoML-for-PICNN}.

\section{Code availability}
All the source codes to reproduce the results can be found on GitHub: \url{https://github.com/wanyunzh/AutoML-for-PICNN}.

\section{Author Contributions}
W.Z. and X.C. designed the Auto-PICNN method. W.Z. and H.S. performed the experiments. W.Z. wrote the manuscript. X.C. supervised the project. All authors reviewed and contributed to revisions of the final manuscript.

\section{Competing interests}
The authors declare no competing interests.


\clearpage
\bibliographystyle{elsarticle-num} 
\bibliography{references}         

\end{document}